\documentclass{elsart}
\usepackage{graphicx}
\usepackage{bm} 
\usepackage{amssymb,amsmath,cite,color}
\journal{Nuclear Physics A}
\newcommand{\degr}{\mbox{$^{\circ}$}}
\newcommand{\bmath}[1]{\mbox{\boldmath ${#1}$}}

\def\fmn#1#2{\mbox{${\textstyle \frac{#1}{#2}}$}}

\newcommand{\dd}{\mbox{\rm d}}
\begin{document}
\begin{frontmatter}
\title{Cross Section and Tensor Analysing Power of the \bmath{dd\to \eta\hspace{1mm}\alpha} Reaction Near Threshold}
\author{The GEM Collaboration:}
\author[c]{A.~Budzanowski},
\author[z]{A.~Chatterjee},
\author[f,w]{R.~Gebel},
\author[e]{P.Hawranek},
\author[d]{R. Jahn},
\author[z]{V.~Jha},
\author[f,w]{K.~Kilian},
\author[c]{S.~Kliczewski},
\author[f,w,m]{Da.~Kirillov},
\author[g]{Di.~Kirillov},
\author[h]{D.~Kolev},
\author[i]{M.~Kravcikova},
\author[e,f,w]{M.~Lesiak},
\author[k]{J.~Lieb},
\author[f,w,m]{H.~Machner}\corauth[cor]{Corresponding author}
\ead{h.machner@fz-juelich.de},
\author[e]{A.~Magiera},
\author[f,w]{R.~Maier},
\author[l]{G.~Martinska},
\author[n]{S.~Nedev},
\author[g]{N.~Piskunov},
\author[f,w]{D.~Prasuhn},
\author[f,w]{D.~Proti\'c},
\author[f,w]{J.~Ritman},
\author[f,w]{P.~von Rossen},
\author[z]{B.~J.~Roy},
\author[g]{I.~Sitnik},
\author[c,d]{R.~Siudak},
\author[h]{R.~Tsenov},
\author[l]{J.~Urban},
\author[f,w,h]{G.~Vankova},
\author[o]{C.~Wilkin}
\address[c]{Institute of Nuclear Physics, Polish Academy of Sciences, Krakow, Poland}
\address[z]{Nuclear Physics Division, BARC, Bombay-400 085, India}
\address[f]{Institut f\"{u}r Kernphysik, Forschungszentrum J\"{u}lich, J\"{u}lich,
Germany}%
\address[w]{J\"{u}lich Centre for Hadron Physics, Forschungszentrum J\"{u}lich, J\"{u}lich, Germany}
\address[e]{Institute of Physics, Jagellonian University, Krak\'{o}w, Poland}
\address[d]{Helmholtz-Institut f\"{u}r Strahlen- und Kernphysik der Universit\"{a}t Bonn, Bonn, Germany}
\address[m]{Fachbereich Physik, Universit\"{a}t Duisburg-Essen, Duisburg, Germany}
\address[g]{Laboratory for High Energies, JINR Dubna, Russia}
\address[h]{Physics Faculty, University of Sofia, Sofia, Bulgaria}
\address[i]{Technical University, Kosice, Kosice, Slovakia}
\address[k]{Physics Department, George Mason University, Fairfax, Virginia, USA}
\address[l]{P.J.~Safarik University, Kosice, Slovakia}
\address[n]{University of Chemical Technology and Metallurgy, Sofia, Bulgaria}
\address[o]{Department of Physics and Astronomy, UCL, London, U.K.}
%
%
\begin{abstract}
The angular distributions of the unpolarised differential cross
section and tensor analysing power $A_{xx}$ of the
$\vec{d}d\to\alpha\,\eta$ reaction have been measured at an excess
energy of 16.6~MeV. The ambiguities in the partial-wave description
of these data are made explicit by using the invariant amplitude
decomposition. This allows the magnitude of the $s$-wave amplitude to
be extracted and compared with results published at lower energies.
In this way, firmer bounds could be obtained on the scattering length
of the $\eta\,\alpha$ system. The results do not, however,
unambiguously prove the existence of a quasi-bound $\eta\,\alpha$
state.
\end{abstract}
\end{frontmatter}
%
%
\section{Introduction}\label{sec:Intro}
\setcounter{equation}{0}

The possibility of the $\eta$ meson forming a quasi-bound state in a
nucleus was first raised by Haider and Liu~\cite{Haider_Liu86}. Such
a state could arise as a consequence of the strongly attractive
$\eta$-nucleon interaction that is driven by the $N^{*}(1535) S_{11}$
resonance. Since the $\eta$ has isospin $I=0$, this leads to
attraction for both the protons and neutrons in the nucleus. A value
of the (complex) $s$-wave $\eta N$ scattering length $a_{\eta
N}=a_r+ia_i \approx (0.28 + 0.19i)$~fm had been
found~\cite{Bhalerao85} on the basis of the existing phase shifts.
Using this value, it was shown that the $\eta$ meson could form a
quasi-bound state with nuclei of mass number $A \ge
10$~\cite{Haider_Liu86}. Other groups found similar results when
starting from this relatively small value of $a_{\eta
N}$~\cite{Hayano99, Garcia-Recio02}.

The likelihood of $\eta$-nucleus quasi-bound states existing would
increase significantly if the $\eta$-nucleon scattering length were
much larger than that assumed by Haider and Liu in 1986. In the
subsequent years, widely differing estimates have been given for
$a_{\eta N}$, with some real parts being as large as 1~fm; see
Ref.~\cite{Haider_Liu02} for a summary. Theoretical studies by
Ueda~\cite{Ueda91} gave some hints that the $\eta$ meson might form a
quasi-bound state with $^{3,4}$He and even with deuteron. The states
could be narrow and lightly bound in few-nucleon systems and
therefore might be observed through their effects above threshold.

According to the Watson-Migdal theory~\cite{Watson52, Migdal55}, when
there is a weak transition to a system where there is a strong final
state interaction (\emph{FSI}), one can factorise the $s$-wave
reaction amplitude, $f_s$, near threshold in the form
\begin{equation}\label{equ:FSI}
f_s=\frac{f_B}{1-iap_\eta}\,,
\end{equation}
where $p_{\eta}$ is the $\eta$ c.m. momentum. The unperturbed
production amplitude $f_B$ is assumed to be slowly varying and is
often taken to be constant in the near-threshold region.

Unitarity demands that the imaginary part of the scattering length be
positive, \emph{i.e.}, $ a_i>0$. In addition, to have binding, there
must be a pole in the negative energy half-plane, which requires
that~\cite{Haider_Liu02}
\begin{equation}\label{equ:Cond2}
\left.{|a_i|}\right/{|a_r|}<1\,.
\end{equation}
Finally, in order that the pole lie on the bound- rather than the
virtual-state plane, one needs also $a_r<0$.

The unexpectedly large near-threshold production amplitude in the
case of the $pd \to \eta^{\,3}$He reaction, as well as its rapid
decrease with rising energy~\cite{Berger,Mayer96}, were interpreted
as evidence for a strong $s$-wave \emph{FSI}, which might be
associated with the formation of a $\eta^{\,3}$He quasi-bound
system~\cite{Wilkin93}. Analogous experimental data on the $dd \to
\eta^{\,4}$He reaction~\cite{Frascaria94,Willis97} show an amplitude
that varied more slowly than in the case of $\eta^{\,3}$He. A
combined analysis to the two data sets within the framework of a
simple optical potential yielded scattering lengths~\cite{Willis97}
\begin{eqnarray}
\nonumber
a(\eta^{\,3}\textrm{He})& =& (-2.3 + 3.2i)~\textrm{fm}\,,\\
a(\eta^{\,4}\textrm{He})& =& (-2.2 + 1.1i)~\textrm{fm}.
\end{eqnarray}
Taken at face value, these suggest that there are poles in both
amplitudes close to threshold but that, whereas the $\eta^{\,4}$He
system might be quasi-bound, the scattering length for $\eta^{\,3}$He
does not satisfy the condition of Eq.~(\ref{equ:Cond2}).
Nevertheless, more recent very refined measurements of the $pd \to
\eta^{\,3}$He differential cross section show even larger
$\eta^{\,3}$He scattering lengths~\cite{Mersmann07,Smyrski07}. These
prove that there is a pole in the complex energy plane within about
1~MeV from threshold and this is confirmed by the energy dependence
of the angular distribution~\cite{Wilkin07}.

In order to isolate the $s$-wave amplitude for $dd \to \eta^{\,4}$He,
it is necessary to measure differential and not merely total cross
sections~\cite{Frascaria94,Willis97}. This has been done by the ANKE
collaboration at an excess energy of $Q=7.7$~MeV~\cite{Wronska05}.
However, as shown there, even differential cross sections are not
completely sufficient for this purpose and the analysis of these and
the earlier data is model dependent. Although there is only one
$s$-wave amplitude, observed deviations from isotropy might be a
purely $p$-wave effect or could result from an interference of the
$s$- with the $d$-wave~\cite{Wronska05}. The only way to decide
between these two solutions is through the determination of a
deuteron analysing power in conjunction with the differential cross
section.

In this paper we present results of a measurement of
$\dd\sigma/\dd\Omega$ and the tensor analysing power $A_{xx}$ at
$Q=16.6$~MeV, which allows us to resolve this ambiguity. The
experiment was carried out using the Big Karl
spectrograph~\cite{drochner98, Bojowald02} and the basic setup is
described in section~\ref{sub:Experim-Setup}. It was carried out in
two stages, first using an unpolarised beam, as discussed in
section~\ref{sub:Unpolar-angular-distrib}. In order to extract an
analysing power, the beam polarisation had to be established and the
way of doing this is explained in section~\ref{polar_setup} and then
used in the measurement of $A_{xx}$ in
section~\ref{sub:Polariz-Cross-Section-Measure}.

The results for the unpolarised differential cross section are
presented in section~\ref{sub:Unpolar-Cross-section} and analysed in
terms of the partial wave and invariant amplitudes~\cite{Wronska05}
in section~\ref{sub:Partial-Wave-Amplitu}. It is shown that, if we
neglect $g$-waves in the $\eta\,\alpha$ system, the magnitude of the
$s$-wave amplitude could be determined from measurements of
differential cross section and $A_{xx}$. The data clearly favour
$s$--$d$ interference as the origin of the angular dependence. The
results are compared in section~\ref{sub:s-wave} with those obtained
at lower energies and together these go a long way in clarifying the
ambiguities regarding the $s$-wave scattering length. Our conclusions
are drawn in section~\ref{sec:summary}.

%
%
\section{Experiment}\label{sec:Experiment}
\setcounter{equation}{0}
%
%
\subsection{Experimental Setup}\label{sub:Experim-Setup}

The GEM collaboration measured the $\vec{d}d \to \alpha\,\eta$
reaction at the COSY-J\"ulich accelerator using vector and tensor
polarised deuteron beams. The beam momentum of 2385.5~MeV/c
corresponds to an excess energy of $Q=16.6$~MeV for this reaction
when an $\eta$-meson mass of $m_{\eta}=547.7$~MeV/$c^2$ is
used~\cite{PDG}. The Big Karl magnetic spectrograph~\cite{drochner98,
Bojowald02} employed for this study is equipped with two sets of
multi-wire drift chambers (MWDC) for position measurement and thus
track reconstruction. Two layers of scintillating hodoscopes, 4.5~m
apart, led to a more accurate time-of-flight measurement than
previously achieved with Big Karl. They also provided the energy-loss
information necessary for particle identification.

\begin{figure}[h]
\begin{center}
\parbox[c]{0.46\textwidth}{
\centering
\includegraphics[width=0.42\textwidth]{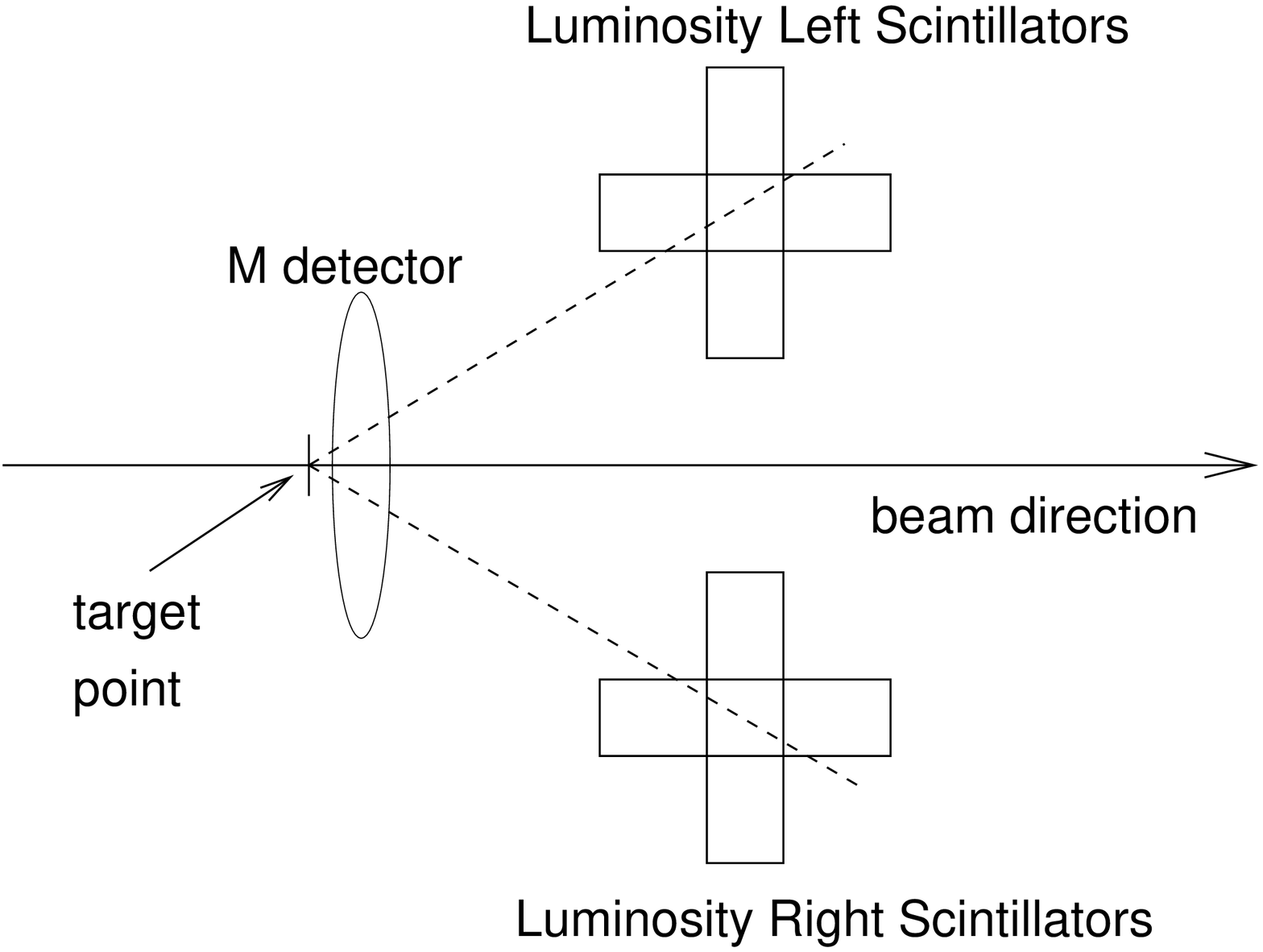}
}
\parbox[c]{0.46\textwidth}{
\centering
\includegraphics[width=0.42\textwidth]{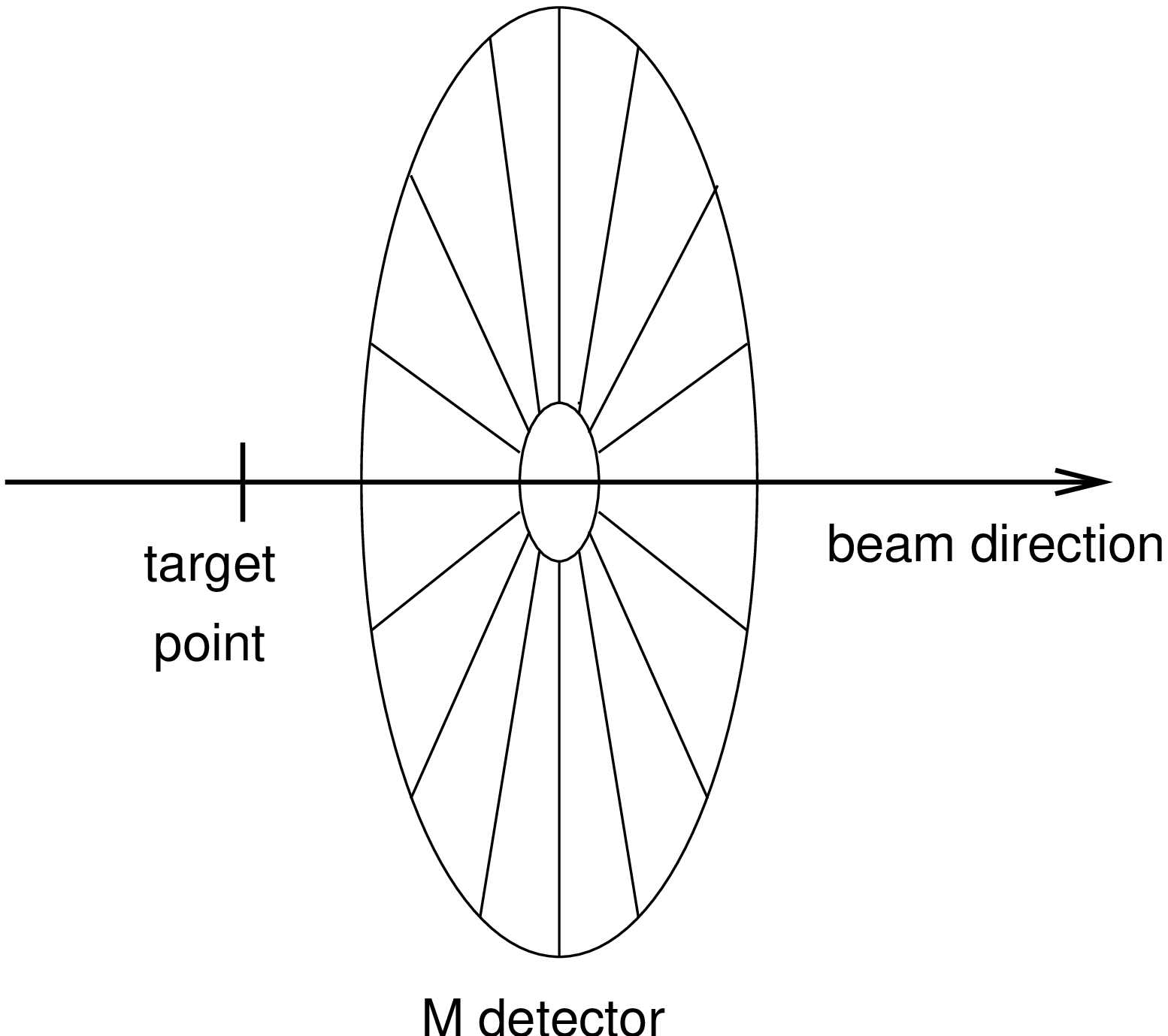}
} \vspace*{8pt} \caption{Schematic layout of the experimental set-up
close to the target area. \textbf{Left}: Positioning of the
luminosity counters. \textbf{Right}: View of the M detector counter
used for monitoring the beam polarisation. }
\label{Fig:m_detector_layout}
\end{center}
\end{figure}

A sketch of the target area is shown in
Fig.~\ref{Fig:m_detector_layout}. The target was a cell of 6~mm
diameter and 4~mm length, which contained liquid deuterium or
hydrogen. The target windows were made from Mylar with a thickness of
less than 1~$\mu$m~\cite{Jaeckle94}. Crossed scintillator paddles,
working in coincidence, were used as luminosity monitors. They were
calibrated by using scattered particles from a beam with reduced
intensity that were directly counted in an exit window in the side
yoke of the dipole $D1$ on the way to the beam dump. The detectors
were therefore named the left and right beam-dump monitors. Whereas
the beam telescope saturates for intensities above about
$10^{\,6}$s$^{-1}$, the luminosity monitors were still far below
saturation at beam intensities of the order of $10^{\,8}$s$^{-1}$.
The luminosities obtained from the two monitors always agreed within
a few percent.

\begin{figure}[!ht]
\begin{center}
\includegraphics[width=9cm]{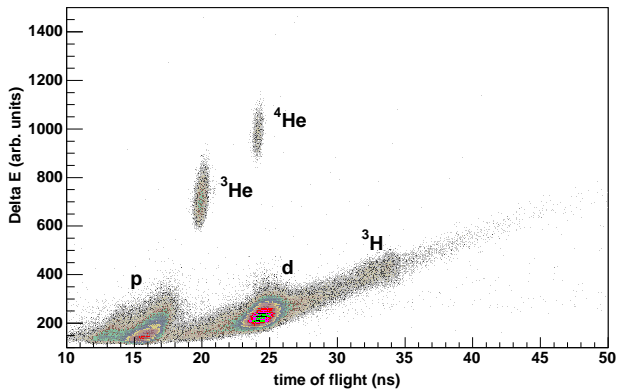}
\caption{Energy losses in the scintillator layers versus the time of
flight between them for the $d(d,X)$ reaction at 2.39~GeV/c. The
islands of $^3$He and $^4$He are well separated from the recoil
particles.} \label{ap_tof}
\end{center}
\end{figure}

The system yielded very good particle separation, as is seen from the
two-dimensional plot of energy loss versus time of flight shown in
Fig.~\ref{ap_tof}.

%
%
\subsection{Unpolarised Angular Distribution}
\label{sub:Unpolar-angular-distrib}

\begin{figure}[!h]
\begin{center}
\includegraphics[width=8 cm]{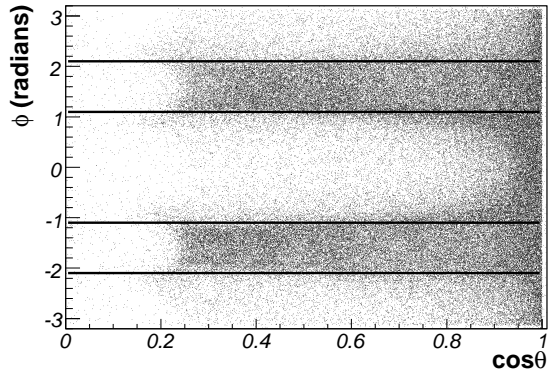}
\caption{Measured $\alpha$-particle events after applying cuts as
function of the polar angle and the cosine of the azimuthal angle in
the center-of-mass frame. The regions around $\phi\approx 0$ and
$\phi\approx \pm\pi$ have almost zero acceptance due to the side yoke
in the first magnetic dipole.} \label{Fig:phi_cos}
\end{center}
\end{figure}

In the first stage of the experiment, we used an unpolarised deuteron
beam incident on the target cell filled with liquid deuterium. The
acceptance of the Big Karl spectrograph for $\alpha$-particles is
limited, not only in momentum, but also geometrically, mainly due to
the side yokes in the first dipole magnet. The consequences of this
for the acceptance are shown in Fig.~\ref{Fig:phi_cos} in terms of
the polar ($\theta$) and azimuthal ($\phi$) angles of the
$\alpha$-particles. The regions with high acceptance that were
retained in the analysis, $\pi/2-0.5\le |\phi|\le \pi/2+0.5$, are
indicated by the solid lines. The simulated acceptance with these
cuts is a smooth function ranging from 0.3 at $\cos\theta =0.4$ to
0.9 at $\cos\theta = 1$.

\begin{figure}[!h]
\begin{center}
\includegraphics[width=8 cm]{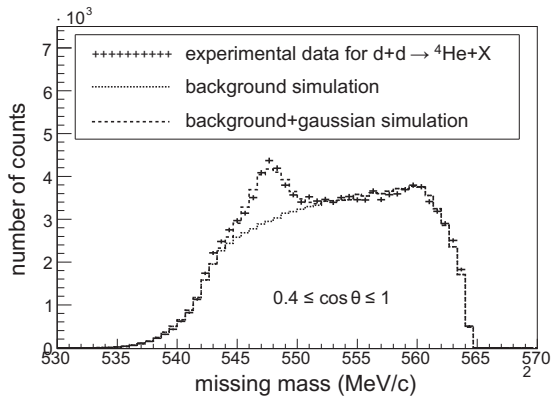}
\caption{Missing mass distribution for the $dd\to\alpha X$ reaction
in the range $0.4 \leq \cos\theta \leq 1$.} \label{Fig:MM}
\end{center}
\end{figure}

After applying cuts, the missing mass of the detected
$\alpha$-particle was evaluated and the event placed in a
$\cos\theta$ bin. The resulting spectrum, summed over all angles is,
shown in Fig.~\ref{Fig:MM}. A clear $\eta$ peak is superimposed on a
background coming from multipion production. The latter was simulated
by calculating the spectral form of two, three and four pion
production in a phase-space model, though the shapes from two and
three pion production were almost identical. The individual strengths
were fitted to the experimental spectrum together with a Gaussian for
the peak and the results shown in Fig.~\ref{Fig:MM}.

\begin{figure}[!h]
\begin{center}
\includegraphics[width=8 cm]{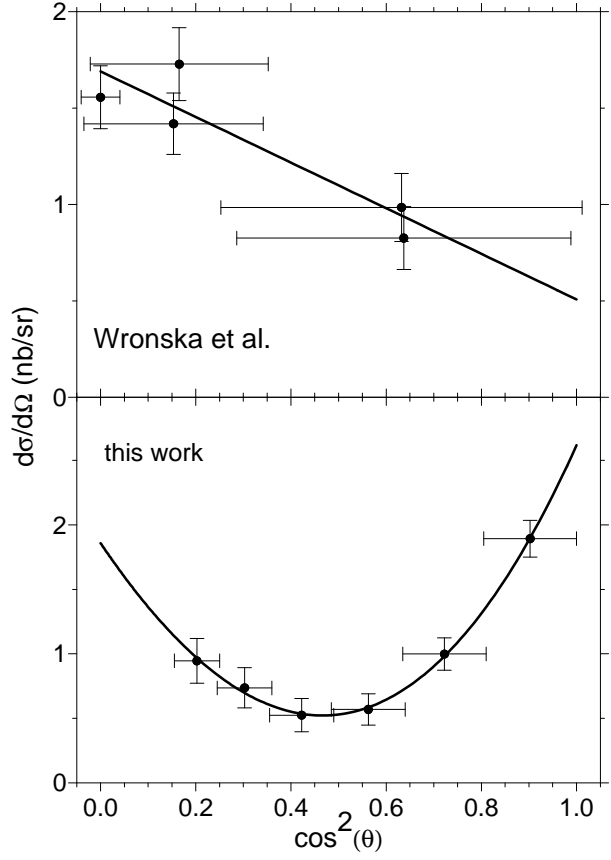}
\caption{Unpolarised $dd\to\alpha\,\eta$ differential cross section.
Since the entrance channel is symmetric, the data are shown in terms
of $\cos^2\theta$. Upper panel: Data from Ref.~\cite{Wronska05} taken
at $Q=7.7$~MeV. The curve is a linear fit in $\cos^2\theta$. Lower
panel: Same as the above but for the present data at $Q=16.6$~MeV.
These data are compared with a quadratic fit in $\cos^2\theta$.}
\label{Fig:Cross_section}
\end{center}
\end{figure}

An analogous analysis was applied to six bins, each of width
$\Delta(\cos\theta)=0.1$. The peak areas were corrected for
acceptance and converted into differential cross sections using the
known luminosity. The resulting differential cross sections are shown
in Fig.~\ref{Fig:Cross_section}.

When the incident deuterons are unpolarised, the entrance channel is
symmetric and a representation of the differential cross section in
terms of $\cos^2\theta$ is appropriate. Already at an excess energy
of 7.7~MeV the ANKE collaboration found deviations from
isotropy~\cite{Wronska05}, with evidence of a forward dip. In
contrast, the present data at $Q=16.6$~MeV show a strong peaking in
the forward direction. Both sets of results are shown in
Fig.~\ref{Fig:Cross_section}. We defer further discussion of these
distributions until section~\ref{sec:Results-Discuss}.

%
%
\subsection{Beam Polarisation}
\label{polar_setup}

The ion source used to generate the polarised deuteron beam is
described in Ref.~\cite{Felden05}. For this experiment, in addition
to the unpolarised beam, two different combinations of vector $p_{z}$
and tensor $p_{zz}$ polarisation were used\footnote{It is customary
to call the quantisation axis ``z'' in the source frame.}:
\begin{eqnarray}\nonumber
p_{z} = -1/3 &\hspace{5mm}\textrm{and}\hspace{5mm}& p_{zz} = +1,\\
\label{notation} p_{z} = -1/3&\hspace{5mm}\textrm{and}\hspace{5mm} &
p_{zz} = -1.
\end{eqnarray}

The vector polarisation $p_{z}$ was measured with a low energy
polarimeter placed in the injection beam line, where the deuteron
energy is about 76~MeV. Using a carbon target, deuteron elastic
scattering was measured with scintillating detectors. The results for
the two polarisation combinations were consistent with the
expectations, with $p_{z} = -0.32 \pm 0.02$ and $p_{z} = -0.33 \pm
0.02$. It has been shown that there is little or no loss of
polarisation in the subsequent acceleration of deuterons at COSY up
to the momentum of interest here~\cite{Chiladze}.

The tensor polarisation of the beam was measured by scattering the
accelerated deuterons from the liquid hydrogen target. The
elastically scattered deuterons were then measured with Big Karl. The
cross section for polarised $\vec{d}p$ elastic scattering in the
backward direction can be expressed as:
\begin{eqnarray}
\left(\frac{d\sigma}{d\Omega}(\theta=180\degr)\right)_{\!\text{pol}}
= \left(\frac{d\sigma}{d\Omega}(\theta=180\degr)
\right)_{\!\text{unpol}}\left[ 1 +
\fmn{1}{2}p_{zz}A_{yy}(\theta=180\degr) \right]\!. \label{dsdo_pzz}
\end{eqnarray}

At strictly 180\degr, there is only one analysing power,
$A_{yy}=A_{xx}$, and this was measured at Saclay~\cite{Arvieux83}
over a broad range of deuteron momenta that covered the 2.39~GeV/c
used here. The points were remeasured with smaller uncertainties by
Punjabi \emph{et al.}~\cite{Punjabi95}, and we made use of their
results.

\begin{figure}[!ht]
\begin{center}
\includegraphics[width=8cm]{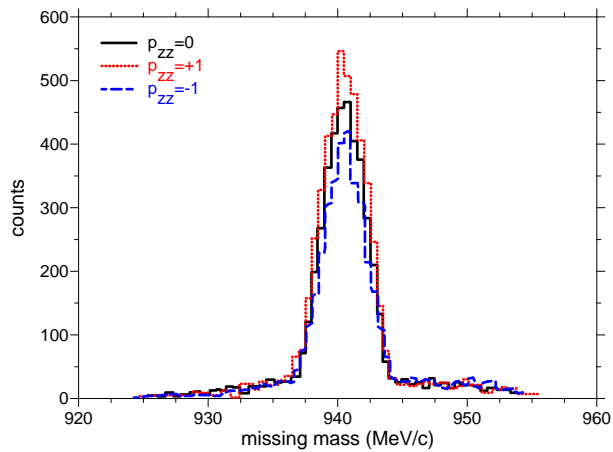}
\caption{{ Missing mass spectrum for the $\vec{d} p \to pd$ reaction
for the unpolarised beam (solid line) and two different tensor
polarisations (dashed and dotted lines).}} \label{mm013}
\end{center}
\end{figure}

The outgoing deuterons were identified using the specific energy
losses and time of flight. In order to eliminate background, the
deuteron missing-mass spectra were analysed. An example of such a
spectrum is shown in Fig.~\ref{mm013} for the $\vec{d}p \to d X$
process for the unpolarised beam and for both states of deuteron
polarisation. In all cases there is a strong proton peak sitting on a
smoothly varying background from which it was straightforward to
extract the numbers of elastically scattered deuterons. The tensor
polarisations summarised in Table~\ref{Tab:pzz_result} were then
determined on the basis of Eq.~(\ref{dsdo_pzz}).

\begin{table}[!ht]
\begin{center}
\caption{Results of the measurements of vector and tensor
polarisations of the deuteron beam. The first errors are statistical
and the second systematic, arising principally from the calibration
polarisations~\cite{Arvieux83,Punjabi95}.\vspace{2mm}
\label{Tab:pzz_result}}
\begin{tabular}{|c|c|c|c|}
\hline
\multicolumn{2}{|c|}{$p_{z}$}&\multicolumn{2}{|c|}{$p_{zz}$}\\
\hline
nominal & measured & nominal & measured\\
                    \hline
$-1/3$ & $-0.33\pm0.02$ & $-1$ & $-0.87 \pm
0.11\pm0.01$ \\
\hline $-1/3$& $-0.32 \pm 0.02$ & $+1$ &
$+0.91 \pm 0.14 \pm0.01$ \\
\hline
\end{tabular}
\vspace{2mm}
\end{center}
\end{table}

%
%
\subsection{Polarised Cross Section Measurement}
\label{sub:Polariz-Cross-Section-Measure} In the subsequent
experiment, we employed the vector and tensor polarised deuteron beam
incident on the deuteron target to measure the analysing power in the
$\vec{d}d\to\alpha\,\eta$ reaction. The following sequence of
polarisation modes was used
\begin{equation}
\mathcal{P}_{+1}\mathcal{P}_{-1} \mathcal{P}_{+1}\mathcal{P}_{-1}
\mathcal{P}_{+1}\mathcal{P}_{-1}\mathcal{P}_{+1}\mathcal{P}_{-1}
\mathcal{P}_{0} \mathcal{P}_{+1}\mathcal{P}_{-1}\cdots ,
\label{pol_sequence}
\end{equation}
where $\mathcal{P}_{i}$, $i$=\(+1,-1\), is a mode with the nominal
values of vector $p_{z}$ and tensor $p_{zz}$ polarisation of the beam
given in Table~\ref{Tab:pzz_result}. It should be noted that, in
contrast to the initial experiment with the unpolarised beam, the
nominally unpolarised mode $\mathcal{P}_{0}$ was prepared by
employing three transitions of the source and so some small residual
polarisation could not be excluded.

In the reaction frame, $\hat{z}$ is taken along the beam direction,
$\hat{y}$ represents the upward normal to the COSY accelerator, and
$\hat{x}=\hat{y}\times\hat{z}$. The polarised
$\vec{d}d\to\alpha\,\eta$ cross section at a production angle
$\theta$ can be written in Cartesian basis as
{\setlength\arraycolsep{2pt}
\begin{eqnarray}\nonumber
\underbrace{\left.\Big( \frac{d\sigma}{d\Omega} (\theta, \phi)
\Big)_{\!\text{pol}}\right/\Big( \frac{d\sigma}{d\Omega} (\theta)
\Big)_{\!\text{unpol}}}_F
&=&\Big[1 +\frac{3}{2}\underbrace{A_{y}(\theta)p_{z}}_H\cos\phi\\
& & \hspace{-5cm}+
\frac{1}{4}\underbrace{p_{zz}(A_{yy}(\theta)+A_{xx}(\theta))}_G +
\frac{1}{4}\underbrace{p_{zz}(A_{yy}(\theta)-A_{xx}(\theta))}_J\cos2\phi
\Big]. \label{dsigma_cartesian_prim}
\end{eqnarray}}

\begin{figure}[!h]
\begin{center}
\includegraphics[width=8cm]{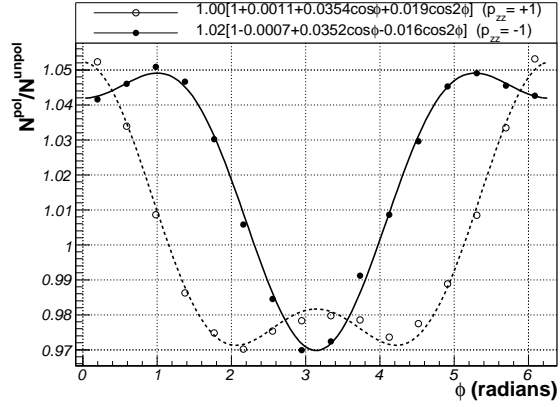}
\caption{Ratio of the counting rate in each M detector segment,
corresponding to a certain azimuthal angle $\phi$, for two different
polarisations of the deuteron beam. The dashed line represents the
fit result for $(p_{z},p_{zz})=(-\frac{1}{3},+1)$, while the solid
line corresponds to $(p_{z},p_{zz})=(-\frac{1}{3},-1)$.}
\label{Fig:momo_dp71}
\end{center}
\end{figure}

The polarisation was inspected for each run by fitting the quantities
$F$, $G$, $H$ and $J$ to the angular distributions on the M detector.
An example of the response of the M detector to different beam
polarisations is shown in Fig.~\ref{Fig:momo_dp71}, together with the
fitted curves.

\begin{figure}[h]
\begin{center}
\parbox[c]{0.48\textwidth}{
\centering
\includegraphics[width=0.48\textwidth]{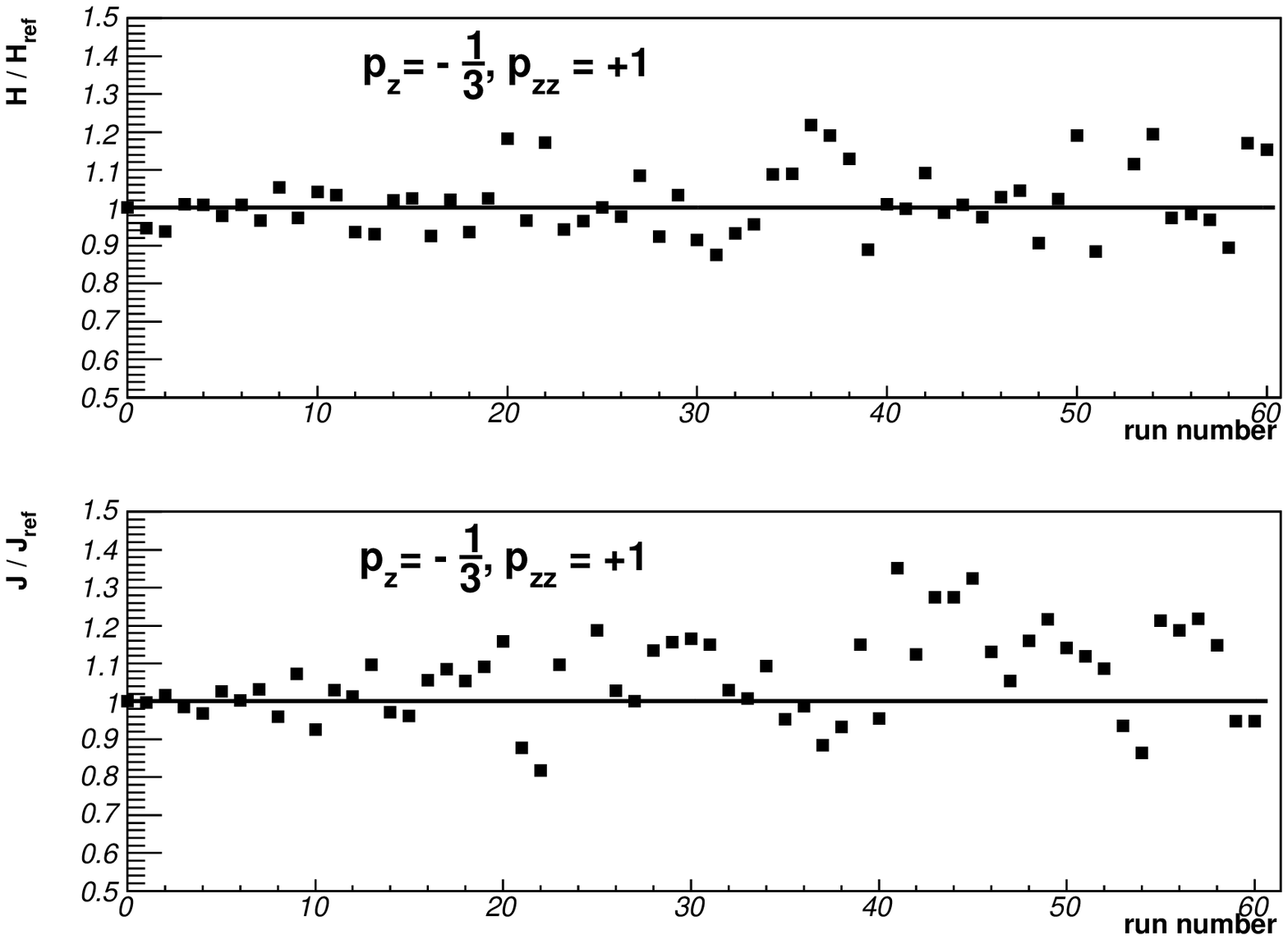}
}
\parbox[c]{0.48\textwidth}{
\centering
\includegraphics[width=0.48\textwidth]{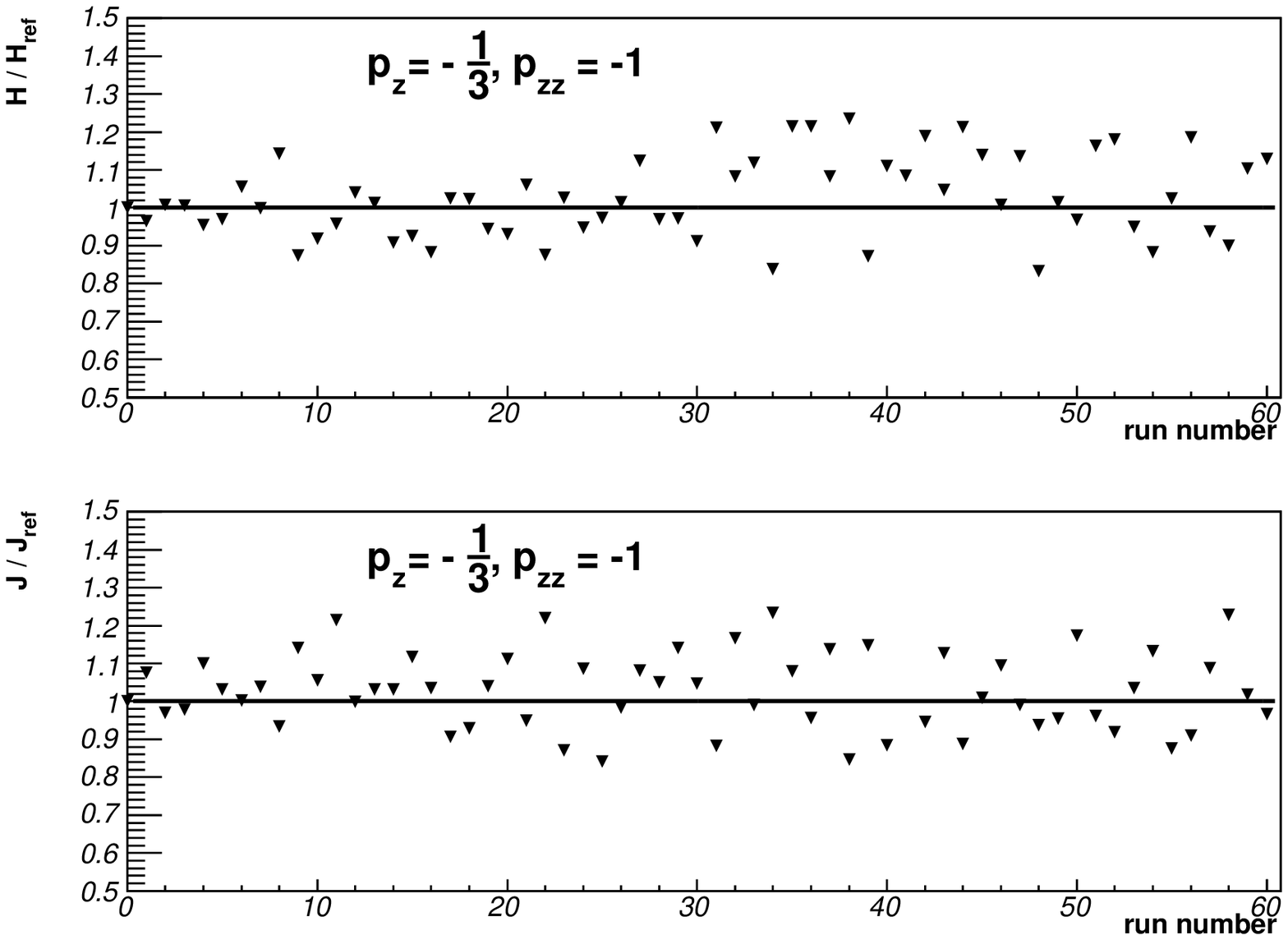}
} \vspace*{8pt} \caption{Comparison of the ratio of the measured
parameters $H$ and $J$ to the reference values $H_{\text{ref}}$ and
$J_{\text{ref}}$ for both polarisation states as functions of the run
number. The $H$ parameter corresponds to the vector polarisation
$p_{z}$, while $J$ is connected with the tensor beam polarisation
$p_{zz}$. Left: Polarisation state with $p_{z}=-\frac{1}{3}$,
$p_{zz}= +1$. Right: Polarisation state with $p_{z}=-\frac{1}{3}$,
$p_{zz}= -1$.} \label{Fig:param_dd}
\end{center}
\end{figure}

The fitted parameters $H$ and $J$ are shown in
Fig.~\ref{Fig:param_dd} relative to the nominal values
$H_{\text{ref}}$ and $J_{\text{ref}}$ for individual run numbers. The
deviations from these values are less than $\pm15\%$. Because of the
smallness of these fluctuations, we retain the nominal values in the
analysis.

The analysing powers could, in principle, be extracted by fitting
Eq.~(\ref{dsigma_cartesian_prim}) directly to the data. However, the
present detector does not have full acceptance (see
Fig.~\ref{Fig:phi_cos}). Furthermore, as is seen from
Eq.~(\ref{pol_sequence}), the unpolarised beam in this part of the
experiment had only 10\% of the statistics and even here there may be
some residual polarisation. We therefore developed a method to
extract the analysing powers without making use of the unpolarised
beam.

As indicated in Fig.~\ref{Fig:phi_cos}, there are two ranges of
azimuthal angle where there is almost complete acceptance,
\emph{i.e.},
\begin{equation}
\fmn{1}{2}(\pi-1) \leq |\phi| \leq \fmn{1}{2}(\pi+1)\,\textrm{rad}.
\label{int_limit_pos}
\end{equation}
For these intervals the mean value of $\cos\phi$ vanishes and that of
$\cos 2\phi$ is given by
\begin{equation}
<\cos 2\phi>=\int_{(\pi-1)/2}^{(\pi+1)/2}\cos 2\phi\,d\phi = -0.84.
\label{dt_cos2phi}
\end{equation}
It therefore becomes clear that $A_y$ could not be measured with the
present layout.

Since $<\cos 2\phi>$ is close to $-1$,
Eq.~(\ref{dsigma_cartesian_prim}) shows that our experiment is
primarily sensitive to the values of $A_{xx}$, with only a small
contamination from $A_{yy}$. We therefore neglect the contribution of
$A_{xx}-A_{yy}$ to the counting rates in these intervals. In this
approximation, the cross section integrated over these intervals of
azimuthal angle becomes simply
\begin{equation}
I=\int_{(\pi-1)/2}^{(\pi+1)/2}\left(\frac{d(\theta,\phi)}{d\Omega}\right)_{\!\text{pol}}
d\phi = \left(\frac{d{\sigma}}{d\Omega}
(\theta)\right)_{\!\text{unpol}} \left[1 + 0.46\,p_{zz}A_{xx}(\theta)
\right], \label{dsigma_rho_Axx}
\end{equation}
where the unpolarised cross section is integrated over the same
$\phi$ range.

Carrying out this procedure for the two polarisation states, we find
that
\begin{equation}
\Delta = \frac{I^+-I_1^-}{I^++I^-} =
\frac{0.23\,A_{xx}\left(p_{zz}^+-p_{zz}^-\right)}{
1+0.23\,A_{xx}\left(p_{zz}^++p_{zz}^-\right)}\cdot
\end{equation}
Using the measured beam polarisations reported in
Table~\ref{Tab:pzz_result}, we can solve to find the value of
$A_{xx}$:
\begin{equation}
A_{xx} = 2.44\,\Delta/(1-0.02\,\Delta). \label{Axx2}
\end{equation}

Since the unpolarised cross section drops out of Eq.~(\ref{Axx2}), it
is sufficient to count the numbers of registered $\alpha\,\eta$
events $N$ per incident beam $n$. The total numbers of beam particles
for the truly unpolarised beam and the two polarisation states are
given in Table~\ref{Tab:beam_lumi}. Uncertainties in the target
thickness and in the acceptance \emph{etc.}\ cancel.

\begin{table}[!h]
\caption{Integrated beam intensities for the unpolarised and
polarised deuteron beams.\vspace{2mm}} \label{Tab:beam_lumi}
\begin{center}
\begin{tabular}{|rr|c|}
\hline \multicolumn{2}{|c|}{Nominal}&Integrated beam\\
$p_{z}$ & $p_{zz}$ &intensity $n$  \\
\hline
0              & 0    & $(12.70 \pm 0.48)\times 10^{13}$ \\
$-\frac{1}{3}$ & $+1$ & $(6.98 \pm 0.62)\times 10^{13}$ \\
$-\frac{1}{3}$ & $-1$ & $(6.19 \pm 0.51)\times 10^{13}$\\
\hline
\end{tabular}
\end{center}
\end{table}

\begin{figure}[!h]
\begin{center}
\includegraphics[width=8 cm]{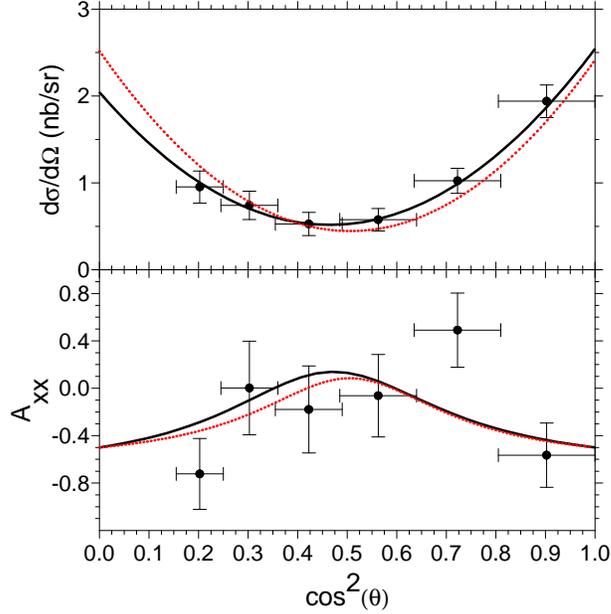}
\caption{Upper panel: Differential cross section for the
$dd\to\alpha\,\eta$ reaction. Lower panel: Analysing power $A_{xx}$.
The solid curves represent a fit with four partial waves; the dotted
curves with invariant amplitudes.} \label{Fig:Partial_wave_fit}
\end{center}
\end{figure}

As is demonstrated later, $A_{xx}$ as well as the unpolarised cross
section are both even functions of $\cos\theta$ and so they are shown
in Fig.~\ref{Fig:Partial_wave_fit} as functions of $\cos^2\theta$.
The large error bars for $A_{xx}$ arise from the fluctuations in the
significant background under the $\eta$ peaks. It should be noted
that the analysing power $A_{xx}$ in the range just above the $\eta$
peak is found to be consistent with zero.

%
%
\section{Results and Discussion}\label{sec:Results-Discuss}
\setcounter{equation}{0}
%
%
\subsection{Unpolarised Cross sections}\label{sub:Unpolar-Cross-section}

In order to deduce the total $dd\to\alpha\,\eta$ cross section, the
differential data were fitted in terms of Legendre polynomials:
\begin{equation}
\frac{\dd\sigma}{\dd\Omega}
=\sum_{l=0}^{\ell_{max}}a_{2l}P_{2l}(\cos\theta).
\end{equation}
The smallest value for $2\ell_{max}$ required to describe our data
was found to be four. The resulting parameters are $a_0=1.27\pm
0.03$~nb/sr, $a_2=-0.29\pm 0.06$~nb/sr, and $a_4=1.65\pm0.07$~nb/sr,
and the corresponding fit is shown in Fig.~\ref{Fig:Cross_section}.
This indicates that there must be at least $d$-wave contributions.
This is to be contrasted to the lower energy ANKE
results~\cite{Wronska05}, where $2\ell_{max}=2$ suffices (see
Fig.~\ref{Fig:Cross_section}). In their case, $a_0 = 1.30\pm
0.18$~nb/sr and $a_2=-0.79\pm 0.19$~nb/sr which, as in our results,
is negative.

\begin{figure}[!h]
\begin{center}
\includegraphics[width=8 cm]{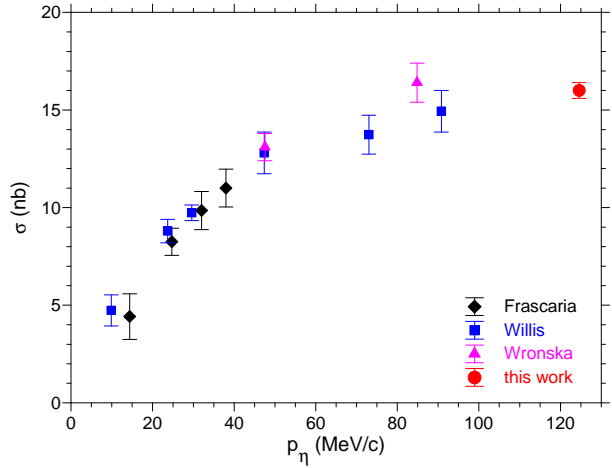}
\caption{Excitation function for the total cross section for the
$dd\to \alpha\,\eta$ reaction. The data are from Frascaria \emph{et
al.}~\cite{Frascaria94}, Willis \emph{et al.}~\cite{Willis97}, and
Wro\'nska \emph{et al.}~\cite{Wronska05}. Only statistical errors are
shown. 
} \label{Fig:Exfu}
\end{center}
\end{figure}

The total cross section deduced from the fit is
\begin{equation}
\sigma = 16.0\pm0.4~\textrm{nb},
\end{equation}
where uncertainties in the target thickness, incident flux, and
acceptance introduce an additional systematic error of $\pm 1.6$~nb.
In Fig.~\ref{Fig:Exfu} this result is compared with existing data,
which are all taken at smaller excess energies.

Willis \emph{et al.}~\cite{Willis97} only extracted the cross section
for helicity $m=\pm 1$, which makes their result for the unpolarised
total cross section model dependent. The values shown in
Fig.~\ref{Fig:Exfu} correspond to the assumption that the $p$-wave
amplitude leading to the $m=0$ cross section is negligible. This is
consistent with our data and we will return to the point in
section~\ref{sub:s-wave}, after discussing the amplitude
decomposition.

%
%
\subsection{Amplitudes and Observables} \label{sub:Partial-Wave-Amplitu}

In order to extract the $s$-wave amplitude for $dd\to\alpha\,\eta$,
we attempted to fit partial wave amplitudes
\begin{equation}
a_{i} = |a_{i}|\,\textrm{e}^{i\phi_{i}}, \label{ampl_ai_def}
\end{equation}
to the present data\footnote{These amplitudes should not be confused
with the scattering length, also denoted by $a$.}. Since the angular
distributions require terms up to $\cos^4\theta$, we consider $s$,
$p$ and (two) $d$ waves. After applying angular momentum and parity
conservation, and taking into account the identical nature of the
incident deuterons, it is seen that there are only four transitions,
which are noted in Table~\ref{Tab:quantum_numbers}.

\begin{table}[!ht]
\caption{Quantum numbers corresponding to the four lowest partial
waves for the $dd\to\alpha\,\eta$ reaction. Here $s_i$ is the total
spin of the initial deuterons with orbital angular momentum $\ell_i$.
Bose symmetry requires that $s_i+\ell_i$ be even. The total angular
momentum $J$ is equal to the $\alpha\,\eta$ orbital angular momentum
$\ell_f$.\vspace{2mm} } \label{Tab:quantum_numbers}
\begin{center}
\begin{tabular}{|c| c c c |c|}
                    \hline
amplitude name &  $s_{i}$ & $\ell_{i}$ & $\ell_f$ & wave name \\
                    \hline
$a_{0}$ & 1 & 1 & 0 & \emph{s} \\
$a_{1}$ & 2 & 2 & 1 & \emph{p} \\
$a_{2}$ & 1 & 1 & 2 & \emph{d} \\
$a_{3}$ & 1 & 3 & 2 & \emph{d} \\
\hline
\end{tabular}
\end{center}
\end{table}

Simonius has shown how to relate such amplitudes to different
possible observables~\cite{Simonius73}. However, when these are
fitted directly to our data it is found that only the magnitude of
the $s$-wave amplitude $|a_0|$ is stable. In contrast, the magnitudes
of the two $d$-wave amplitudes $|a_2|$ and $|a_3|$ are completely
correlated. To see the origin of these effects, it is simplest to
consider the relation of the partial waves to the invariant amplitudes
used by Wro\'nska \emph{et al.}~\cite{Wronska05}, which we briefly
summarise here.

Due to the identical nature of the incident deuterons, only three
independent scalar amplitudes are necessary to describe the spin
dependence of the reaction. If we let the incident deuteron cms
momentum be $\vec{p}_d$ and that of the $\eta$ be $\vec{p}_{\eta}$,
then one choice for the structure of the transition matrix
$\mathcal{M}$ is%
\begin{eqnarray}
\nonumber \mathcal{M}&=&A({\vec{\epsilon}}_1\times{\vec{\epsilon}}_2)
\cdot{\hat{p}_d} +B({\vec{\epsilon}}_1\times{\vec{\epsilon}}_2)\cdot
\left[\hat{p}_d\times(\hat{p}_{\eta}\times\hat{p}_d)\right]
(\hat{p}_{\eta}\cdot{\hat{p}_d})\\
&&\hspace{-2mm}+C\left[({\vec{\epsilon}}_1\cdot{\hat{p}_d})\,
{\vec{\epsilon}}_2\!\cdot\!(\hat{p}_{\eta}\times\hat{p}_d)
+({\vec{\epsilon}}_2\cdot{\hat{p}_d})\,
{\vec{\epsilon}}_1\!\cdot\!(\hat{p}_{\eta}\times\hat{p}_d) \right],
\label{A1}
\end{eqnarray}
where the ${\vec{\epsilon}}_i$ are the polarisation vectors of the
two deuterons.

If we retain up to $d$ waves in the final system, then $B$ and $C$
have no angular dependence while that of the amplitude $A$ can be
written as
\begin{equation}
A(\theta) = A_0 + A_2\,P_2(\cos\theta)\,.
\end{equation}

The observables in a spherical basis were related to these amplitudes
in Ref.~\cite{Wronska05} but they can be easily converted to the
Cartesian observables used here~\cite{Ohlsen}. Our experiment only
yielded data on the unpolarised cross section and tensor analysing
power $A_{xx}$ and these are given in terms of the invariant amplitudes by
\begin{eqnarray}
\nonumber \left(1 - A_{xx}\right)\frac{\dd\sigma}{\dd\Omega} &=&
\frac{p_{\eta}}{p_d}\left(|A_0|^2
+2\emph{Re}(A_0A_2^*)P_2(\cos\theta)+|A_2|^2\left(P_2(\cos\theta)\right)^2\right)\,,\\
\label{comb1}\\
\left(1 + 2 A_{xx}\right)\frac{\dd\sigma}{\dd\Omega}
&=&2\frac{p_{\eta}}{p_d}\left(|B|^2\sin^2\theta\cos^2\theta+|C|^2\sin^2\theta\right)\,,
\label{comb2}
\end{eqnarray}
where the results have been expressed in terms of convenient linear
combinations. From these it is seen that both the cross section and
$A_{xx}$ are even functions of $\cos\theta$, a result that has
already been used in our analysis and presentation. The data could
therefore \emph{in principle} fix the magnitudes of the amplitudes
$A_0$, $A_2$, $B$, and $C$, and the interference between $A_0$ and
$A_2$, while being completely insensitive to all the other phases.
Furthermore, the linear combinations of Eqs.~(\ref{comb1}) and
(\ref{comb2}) show that the fitting of $|B|$ and $|C|$ is decoupled
from that of $A$. The parameters resulting from fitting the data in
this basis are given in Table~\ref{tab:MA_Fits}, with the fit curves
being shown in Fig.~\ref{Fig:Partial_wave_fit}. From this it is seen
that the $A$ amplitudes are dominant, with  $C$ being consistent with
zero within error bars. If $|B|$ also vanished, it would follow from
Eq.~(\ref{comb2}) that $A_{xx}=-\frac{1}{2}$ for all angles. On the
other hand, even the small contribution from the $B$ term changes the
angular dependence of $A_{xx}$, as is evident in
Fig.~\ref{Fig:Partial_wave_fit}.

\begin{table}[!h]
\caption{Fit results of invariant amplitudes of Eqs.~(\ref{comb1}) and
(\ref{comb2}) to the present data. Since $|C|$ was found to be zero
within error bars, it was put exactly to
zero.\vspace{2mm}}\label{tab:MA_Fits}
\begin{center}
\begin{tabular}{|c|c|}
\hline
fit parameter &value \\
\hline
$|A_0|^2$ & $\phantom{-1}6.6\pm1.7$ \\
$2\textit{Re}\,(A_0^*A_2)$ & $-25.0\pm9.5$ \\
$|A_2|^2$ & $\phantom{-1}48.4\pm14.5$ \\
$|B|^2$ & $\phantom{-1}9.3\pm5.1$ \\
$|C|^2$ & $\phantom{-1}0$ \\
\hline
\end{tabular}
\end{center}
\end{table}

The Wro\'nska \emph{et al.}\ amplitudes can be easily expressed in
terms of those of the partial waves $a_i$ of
Table~\ref{Tab:quantum_numbers}. Explicitly
\begin{eqnarray}
\label{relationship}%
\nonumber
\sqrt{\frac{p_{\eta}}{p_d}}\sqrt{4\pi}\,A_0 &=&\frac{1}{3\sqrt{2}}\,a_{0}\,,\\
\nonumber \sqrt{\frac{p_{\eta}}{p_d}}\sqrt{4\pi}\,A_2
&=&\sqrt{5}\left[\frac{1}{\sqrt{14}}\,a_{3}-\frac{1}{3}\,a_{2}\right],\\
\nonumber
\sqrt{\frac{p_{\eta}}{p_d}}\sqrt{4\pi}\,B&=&\frac{\sqrt{5}}{2}
\left[a_{2}+\sqrt{\frac{2}{7}}\,a_{3}\right],\\
\sqrt{\frac{p_{\eta}}{p_d}}\sqrt{4\pi}\,C&=&\frac{1}{2}\sqrt{\frac{3}{5}}\,a_{1}\,.
\end{eqnarray}

From the above discussion, it is seen that if $\ell_\eta=4$ waves are
neglected, then the data might determine the absolute magnitudes of
the $\ell_\eta=0$ and $\ell_\eta=1$ amplitude, \emph{viz.}\ $|a_{0}|$
and $|a_{1}|$. On the other hand, only two linear combinations of the
$d$-wave amplitudes $|a_{2}+2a_{3}/\sqrt{14}|^2$ and
$|a_{2}-3a_{3}/\sqrt{14}|^2$ could fixed by the data. Thus the
differential cross section and tensor analysing power $A_{xx}$ are
together only sensitive to $7|a_{2}|^2+3|a_{3}|^2$ rather than the
individual magnitudes. This accounts for the complete correlation
found in the direct fitting of the partial wave amplitudes to the
data.

%
%
\subsection{The s-Wave Amplitude}\label{sub:s-wave}

We have seen that, provided $g$-waves are neglected, the present data
allow us to extract the magnitude of the $s$-wave amplitude $|a_0|$.
From this we obtain a spin-averaged square of the $s$-wave amplitude,
$|f_s|^2$ through
\begin{equation}\label{Eq:f_s}
\frac{\dd\sigma_s}{\dd\Omega}=\frac{p_\eta}{p_d}|f_s|^2 =
\frac{2p_\eta}{3p_d}|A_0|^2= \frac{1}{27} \frac{1}{4\pi}|a_0|^2,.
\end{equation}
Using the value given in Table~\ref{tab:MA_Fits}, we find that
$|f_s|^2 = 4.4\pm 1.1$ nb/sr.

Close to threshold, where $s$-waves dominate, $|f_s|^2$ can be
extracted from the total cross section $\sigma$ via
\begin{equation}\label{equ:threshold}
|f_s|^2= \frac{p_d}{p_\eta}\frac{\sigma}{4\pi}\,,
\end{equation}
since the contribution from $s$-$d$ interference drops out.

Equation~(\ref{equ:threshold}) has been used by Willis \emph{et
al.}~\cite{Willis97} to extract $|f_s|$ but there are two important
considerations here. Our data show that the $B$ and $C$ amplitudes
play only a minor role in the observables and this is likely to be
even more true at lower energies. Hence to a good approximation the
$m=0$ cross section can be neglected and the total cross section
should be two thirds of their $m=\pm1$ value. This result has already
been used when plotting the data in Fig.~\ref{Fig:Cross_section}.

Furthermore, although the SPESIII detector had complete acceptance
for the $\alpha$ particles from $\eta$ production, the full polar
angle could not be reconstructed and the authors assumed isotropic
distributions~\cite{Willis97}. While this is a good approximation
close to threshold, it is not valid at their two highest energies, as
is evident when comparing with the data of Ref.~\cite{Wronska05}.
There will therefore be $d$-wave contributions which must be
subtracted in Eq. (\ref{Eq:f_s}) before extracting $|f_s|^2$. To do this, we assume that
the $d$-wave amplitudes $A_2$ and $B$ vary with a threshold factor of
$p_\eta^2$, where the normalisation is fixed by our results given in
Table~\ref{tab:MA_Fits}.

\begin{figure}
\begin{center}
\includegraphics[width=8 cm]{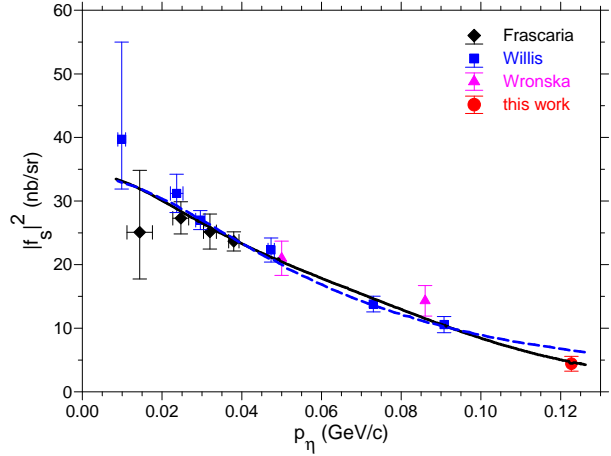}
\caption{The world data for spin-averaged square of the magnitude the
$s$-wave amplitude for $dd \to \eta^{\,4}$He as a function of the
$\eta$ cm momentum. The dashed curve shows the best fit
Eq.~(\ref{Eq:Fermi-exp}) with parameters given in the text. The solid
curve is a fit with Eq.~(\ref{Eq:effective_range}).} \label{Fig:F_s2}
\end{center}
\end{figure}

This procedure results $|f_s|^2=13.8\pm1.2$~nb/sr at
$p_{\eta}=73$~MeV/$c$ and $10.6\pm1.3$~nb/sr at $p_{\eta}=91$~MeV/$c$
for the Willis data~\cite{Willis97} and $14.3\pm2.4$~nb/sr for the
Wro\'{n}ska measurement at 86\,MeV/c~\cite{Wronska05}. All the values
of $|f_s|^2$ are then shown in Fig.~\ref{Fig:F_s2}.

Willis \emph{et al.}~\cite{Willis97} did a combined optical-model fit
to all the near-threshold $pd\to\eta^{\,3}$He and $dd\to\eta
^{\,4}$He data as discussed in the introduction, assuming that only s-wave production occurred, and
found a value of the $\eta\,\alpha$ scattering length of $a = (-2.2 +
i1.1)$~fm. Given the extended $dd\to \eta^{\,4}$He data set now
available, we have attempted a direct fit in the scattering-length
approximation
\begin{equation}\label{Eq:Fermi-exp}
|f_s|^2=\frac{|f_B|^2}{1+p_\eta^2(a_r^2+a_i^2)+2p_\eta a_i}
\end{equation}
The best fit shown in Fig.~\ref{Fig:F_s2} is obtained for
$|a_r|=3.1\pm 0.5$~fm, $a_i=0\pm0.5$~fm and $f_B^2=34\pm 1$~nb/sr,
corresponding to a quasi-bound or virtual state with $|Q_0|\approx
4$~MeV. These above-threshold data are, of course, insensitive to the
sign of $a_r$ so that they could never tell whether the system is
quasi-bound or virtual. The argument given in Ref.~\cite{Willis97} is
that, since $\eta^{\,4}$He is more likely to be bound than
$\eta^{\,3}$He, the fact that $|Q_0|$ is smaller for
$\eta^{\,3}$He~\cite{Mayer96,Mersmann07,Smyrski07}, suggests that
$\eta^{\,4}$He is indeed quasi-bound.

To try to explore the systematic uncertainties, one can include also
an effective range term $r$, as has been done for the near-threshold
$dp \to \eta^{\,3}$He data~\cite{Mersmann07} and fit
\begin{equation}\label{Eq:effective_range}
f_s=\frac{f_B}{1-ip_\eta a+\frac{1}{2}arp_\eta^2}.
\end{equation}
However, due to the strong coupling between the fit parameters, the error
bars become exceedingly large, $a_r=6.2\pm 1.9$~fm and $a_i=0.0.001\pm
6.5$~fm. This fit is also shown in Fig.~\ref{Fig:F_s2}. The large
errors here do not allow a decisive answer on the position of the
$\eta\,\alpha$ pole.
%
%
\section{Summary}\label{sec:summary}
\setcounter{equation}{0}

We have measured the differential cross section and tensor analysing
power $A_{xx}$ of the $dd\to \alpha\,\eta$ reaction at an excess
energy of 16.6~MeV. The recoiling $\alpha$-particles were measured in
a magnetic spectrograph and the $\eta$ mesons identified through a
missing-mass technique. The biggest uncertainty in the method arises
from the large background coming from multipion production. Despite
this, an angular distribution and a total cross section could be
given.

The angular distribution of the analysing power $A_{xx}$ was measured
with a tensor polarised beam. Since this observable required
measurements with two different spin modes, the uncertainty due to
the background subtraction is even larger. Nevertheless, the values
thus obtained showed that $p$-wave amplitude was very small so that
deviations from isotropy must come primarily from $s$--$d$
interference~\cite{Willis97, Wronska05}.

Combining these results with the unpolarised cross section data
allowed the spin-averaged square of the $s$-wave amplitude $|f_s|^2$
to be extracted with reasonable error bars. Assuming that the
$d$-wave production cross section varied like $Q^2$, values of
$|f_s|^2$ were also deduced from earlier measurements such that there
are now 13 values for $Q\leq 16.6$~MeV. Fitting the momentum
dependence in the scattering length
approximation~\cite{Watson52,Migdal55} gives $|a_r|=3.1\pm 0.5$~fm
and $a_i=0.0\pm 0.5$~fm. This suggests that there is a singularity in
the complex plane close to threshold but its position is far from
clear and the uncertainty grows if a fit is attempted with an
effective range as well as a scattering length. Further data are
clearly needed!

%
%
\section*{Acknowledgements}\label{sec:Acknowl}
We are grateful to the COSY crew for providing quality deuteron
beams. Discussions with A.~Wro\'{n}ska were helpful. We appreciate
the support received from the European community research
infrastructure activity under the FP6 ``Structuring the European
Research Area'' programme, contract no.\ RII3-CT-2004-506078, from
the Indo-German bilateral agreement, from the Research Centre
J\"{u}lich (FFE), and from GAS Slovakia (1/4010/07).
%
%

\end{document}